\documentclass[%
 reprint,
 amsmath,amssymb,
 aps,
]{revtex4-2}

\usepackage{graphicx}
\usepackage{dcolumn}
\usepackage{bm}

\begin{document}

\title{A Spontaneous Symmetry Breaking Machine\\
-A Theory for a Novel Type of Spontaneous Symmetry Breaking in a Unique Dissipative System and one Application-}

\author{Toshiya Sato}
\email[Toshiya Sato]{:tshy.sato@ntt.com}

\affiliation{NTT Device Technology Laboratories, NTT Corporation, 3-1, Morinosato Wakamiya, Atsugi-shi, Kanagawa, Japan 243-0198}

\date{\today}

\begin{abstract} 
We focus on an interesting dissipative system found in a photonics system. In this dissipative system, we theoretically identified that robust causality is generated and as a result, it becomes possible to produce behavior that can be understood as SSB, and, we experimentally demonstrated this finding. Furthermore, we theoretically demonstrated that by combining such dissipative systems as fundamental elements and establishing a certain relationship between them through optical interference, it is possible to create a unique system that generates complex SSB as a whole. This unique SSB can be understood as having a duality with the model of the creation of many-body-like system (MBLS), and by using the correspondence between the many-body-like system and the Ising model, it holds promise as an alternative computational resource for solving combinatorial optimization problems.
\end{abstract}

\maketitle
\section{Introduction}
Symmetry breaking (SB) lurks behind various phenomena and is simultaneously both a universal and a basic concept that allows us to understand nature.
As Anderson argued in his scientifically inspiring article, "More is Different" \cite{Anderson}, there is an inevitability about the hierarchical structure of science from the viewpoint of emergence when SB occurs.
A well-known example of SB (type I) is spontaneous magnetization in the thermodynamic limit \cite{Onsager}, which can be understood as the appearance of an ordered spin state in a many-body equilibrium system.
And, it is known that it is difficult to reconstruct the most stable state starting from knowledge of interactions, and that the Ising model system, which simplifies this phenomenon, is an NP-hard problem \cite{Barahona}.

 Other examples of SB (type II) can be found in dissipative systems.
In that area, concepts such as dissipative structure \cite{Prigogine} (e.g. Benard convection \cite{Getling}) and synergetics \cite{Haken} (e.g. lasers) have been proposed, and the emergence of a dynamical macroscopic order has attracted attention.
Another example of SB (type III) can be found in the standard model of elementary particle physics.
The Higgs mechanism, i.e., the mechanism for mass generation of W and Z bosons, was proposed \cite{Higgs} following Nambu's introduction \cite{Nambu} of the concept of spontaneous symmetry breaking (SSB) in superconductivity \cite{BCS, Bogoliubov} into particle physics.
Note that SSB in the Higgs mechanism has two features that distinguish it from types-I and -II SSB.
Firstly, the focus is not on {\it the macroscopic properties derived from the particles that make up the system under observation} but on {\it the global symmetry of the Higgs field itself.}
The other thing that appears through SSB is {\it the microscopic properties of individual elementary particles} known as mass.
Numerous other types of symmetry breaking have been reported, but we attempted to understand the complex symmetry breaking introduced in later sections by drawing upon insights and perspectives gained from studying the three types described here.

In recent years, combinatorial optimization problems (COPs) have received increasing interest in an extremely wide range of fields from social life to academia (logistics, financial management, drug discovery, machine learning, etc.).
Meanwhile, various approaches have been investigated to solve the "Ising problem" \cite{Barahona, Hopfield, Kirkpatrick, Hinton, HopfieldT, Lucas}, which can be mapped to COPs that are NP-hard.
In particular, Ising machines such as quantum annealers (a type of quantum computer)  \cite{Nishimori, Johnson} and coherent Ising machines (based on laser oscillation phenomena) \cite{Yamamoto, Takata}, which use natural phenomena to find the ground state of an Ising model system (IMS), have attracted much attention.

This paper discusses the unique dissipative systems observed in photonic systems and describes the following three points: (1) We theoretically identified and experimentally verified that under specific conditions in the dissipative systems we focus on, robust causality can be generated and by further narrowing the conditions based on this causality, SSB phenomena can be induced. (2) We theoretically demonstrate that by combining multiple such dissipative systems and implementing specific interrelationships between them, it is possible to construct a composite system capable of inducing unique and complex SSB phenomena. (3) Through numerical simulations, we find that this complex SSB phenomenon holds promise as a computational resource for combinatorial optimization problems.

\begin{figure}[htbp]
\centering\includegraphics[width=8cm]{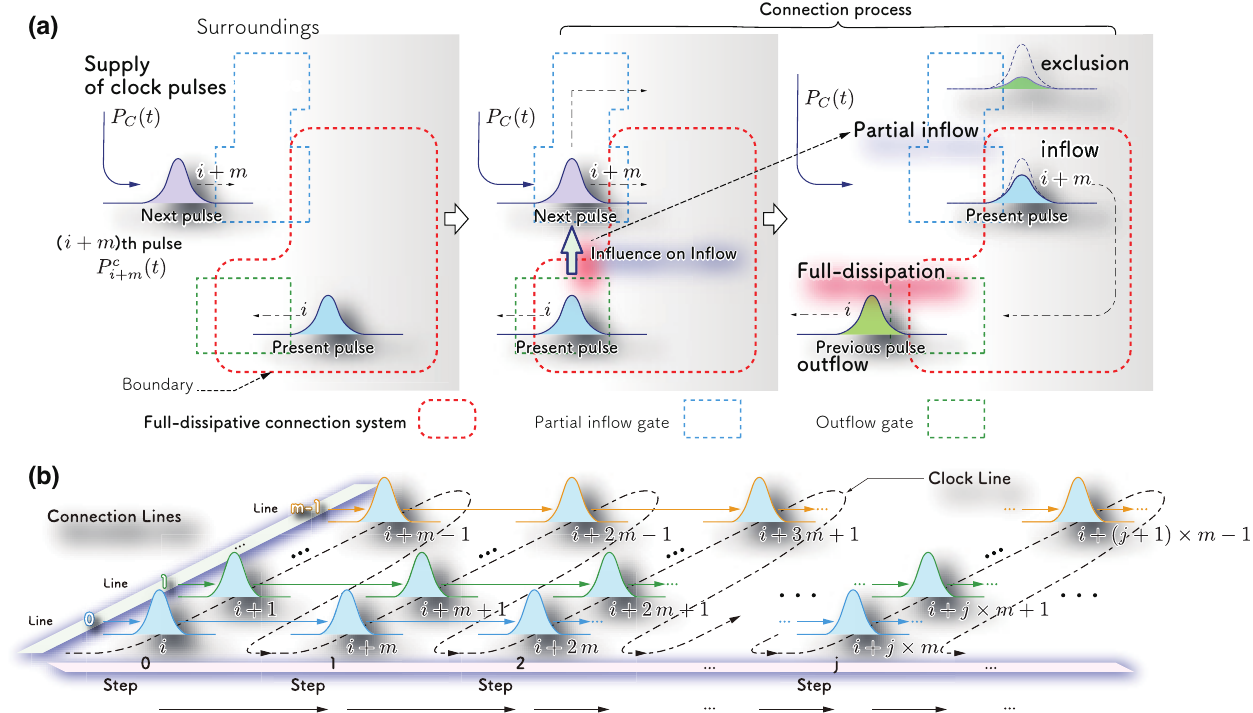}
 \caption{Schematic of a dissipative system we focused on.
(a) This system maintains a state composed of any one of inflow pulses derived from clock pulses. (b) Clock pulses form $m$ of independent dissipative systems (corresponding to each of connection lines).}
 \end{figure}

\section{Dissipative causality in full-dissipative connection system}
First, we will explain a dissipative system that plays a role as a basic element, and then we will look at how robust causality appears in this system.

We consider a dissipative system with a boundary as shown in Fig.1.
The system receives optical clock pulses (OCPs), and the inflow pulse derived from the $i$-th OCP affects the inflow of the $(i + m)$-th OCP (i.e., the next inflow pulse to the system) and then flows out of the system completely.
As a result, the clock pulses form $m \,(\in N)$ elementary (and foundational) dissipative systems, which correspond to the connection lines in Fig.1(b).
Henceforth, we specifically refer to such a dissipative system as a full dissipative connection system (FDCS).
The FDCS can be physically implemented with the construction of an optical experiment system using photonics technology as shown in Fig.2.

\begin{figure}[b]
\centering\includegraphics[width=7cm]{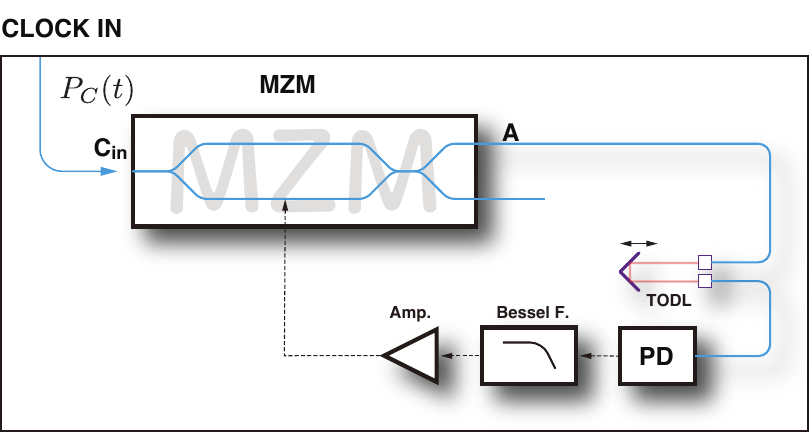}
 \caption{Schematic of a circuit realizing a full-dissipative connection systems (FDCSs). Polarized optical coherent clock pulses (OCPs) $P_{C}(t)$ are supplied from an external input port $\bf{C_{\rm in}}$ to generate the FDCSs and 1$\times $2 MZM is used as \it a partial inflow gate\rm. Number of the FDCSs and timing of electrical driving pulses to the MZM are tuned by both repetition interval of the OCPs $\Delta t$ and delay time by an optical tunable delay line (OTDL). Furthermore, width of the electrical driving pulses to the MZM are stretched sufficiently wider than the OCP width. }
 \end{figure}

Through an optical clock input port (${\bf C}_{\rm in}$), a 1$\times $2 Mach-Zehnder interferometer-type waveguide LiNbO3 intensity modulator (MZM) receives polarized, optically coherent clock pulses $P_{C}(t)$ with peak power $P_{0}$, pulse width $\sigma_{\mbox{\rm \scriptsize pw}}$, and repetition interval $\Delta t$.
This MZM serves as a variable inflow gate for the FDCS.
For the $i$-th pulse $P^{c}_{i}(t)$ that makes up $P_{C}(t)$, the pulse envelope can be expressed using the normalized amplitude $A_{0}$ as follows:

 \begin{eqnarray}
 P^{c}_{i}(t)
& = &
\frac{P_{0}\,\mid A_{0} \mid^{2}}{ \sqrt{ 2 \pi \sigma^{2} } }\,
\exp( - \frac{( t - i \times \Delta t)^{2}}{ 2 \sigma^{2} } )
\\
P_{C}(t)
& = &
\sum_{i}\,
 P^{c}_{i}(t), \,\,\,\,\, i \in N
\end{eqnarray}

The output from one of the MZM's output ports ($ {\bf A}$ in Fig. 2) flows into the FDCS of interest.
This inflow derived from $P^{c}_{i}(t)$ is input into a photodiode (PD) through a tunable optical delay line and converted to an electrical pulse (EP$_{i}$).
After sufficient stretching by a Bessel filter, EP$_{i}$ drives the MZM and is terminated.
Here, the tunable optical delay line matches the arrival time of the peak in EP$_{i}$ at the MZM to the arrival time of $P^{c}_{i+m}(t)$, where $1 < m$.
Then, $P^{c}_{i+m}(t)$ is divided between the outputs $ {\bf A}$ and $ {\overline{\bf A}}$ through correspondence with the Pockels effect induced by the driving pulse EP$_{i}$, and the output from ${\bf A}$  flows into the FDCS.
Here, the dissipation (full dissipation) of all optical pulse energy derived from $P^{c}_{i}(t)$ and the inflow derived from $P^{c}_{i+m}(t)$ are a series of phenomena.
Now, if the pulse width $\tau_{\mbox{\rm \scriptsize pw}}$ of EP$_{i}$ is stretched to be sufficiently wider than the pulse width of $P^{c}_{i}(t)$, then EP$_{i}$ can be considered constant with respect to $P^{c}_{i+m}(t)$.
Moreover, if we can tune the repetition interval of $P_{C}(t)$ to be sufficiently longer than the pulse width $\tau_{\mbox{\rm \scriptsize pw}}$, then we can treat the rises and falls of EP$_{i}$ as step-like changes that have no effect except in $P^{c}_{i+m}(t)$.
Therefore, when the condition $\sigma_{\mbox{\rm \scriptsize pw}} \ll \tau_{\mbox{\rm \scriptsize pw}} \ll \Delta t$ is satisfied, the $(i+m)$-th optical inflow into the FDCS can be expressed as follows:

\begin{eqnarray}
\phi_{i+m} \,  P^{c}_{i+m}(t)
& = &
\sin^{2}( \, \Delta\theta_{i+m}( \,\phi_{i} \,  P^{c}_{i}(t) \, )  + \theta_{B} \, ) \, P_{C}(t)
\nonumber\\
& = &
\sin^{2}( \frac{\gamma}{2} \, \phi_{i} + \theta_{B} \,) \, G( t, \tau_{pw}, (i+m)\Delta t )  P_{C}(t)
\nonumber\\
& = &
\sin^{2}( \frac{\gamma}{2} \, \phi_{i} +  \theta_{B} \, ) \,  P^{c}_{i+m}(t)
\end{eqnarray}

Here,  $\Delta\theta_{i+m}(\phi_{i} \, P^{c}_{i}(t) )$ is the phase difference induced by EP$_{i}$ between the MZM's interference arms; $\gamma, \phi_{i+m}$, and $\theta_{B}$ are the efficiency induced by the phase difference, the transmittance by EP$_{i}$ to the port $ {\bf A}$ for $P^{c}_{i+m}(t)$ and the static phase condition of MZM, respectively; and $G( t, \tau_{pw}, (i+m)\Delta t )$ is the gate effect on $P_{C}(t)$, which can be expressed via the Heaviside step function.
From the above, it can be understood that the FDCS can be described by the following iterative equation, and that $m$ independent and robust causalities are established\cite{Comments003}.

\begin{eqnarray}
\phi_{i+m} 
& = &
\sin^{2}(\frac{\gamma}{2} \, \phi_{i}  +  \theta_{B} ) \,\,\,  \vert_{P_{i+m}^{c}(t) \ne 0}
\end{eqnarray}

There is no doubt that space and time are the fundamental parameters of FDCS, but what is important to note here is that the robust {\it dissipative causality} that emerges in the FDCS system is described without these fundamental parameters.
For another example of argument based on {\it causality} divorced from the framework of space-time, we can cite theoretical studies in the {\it Planck-scale realm,} i.e., the {\it quantum gravity realm}.
In that area, in order to overcome the major difficulty of being unable to discuss the issue based on the assumption of space-time, challenging theoretical research in which causality (causal structure) plays an indispensable role is being carried out in various ways, including causal sets \cite{Sorkin, Henson}, causal dynamic triangulation \cite{Ambjorn}, and the IIB matrix model \cite{Tsuchiya}.
What we want to emphasize here is that {\it dissipative causality} can be experimentally observed and confirmed by monitoring the inflow from $P^{c}_{i+m}(t)$, and that the phenomenon observed there is an important example of a phenomenon that is described not by space-time but by causality.

\section{Novel type of SSB in dissipative causality}

In the previous section, we looked in detail at the conditions under which FDCS produces dissipative causality.
This dissipative causality has two degrees of freedom remaining, and FDCS exhibits extremely diverse behavior by changing these values.
Here, we focus on dissipative causalities under the condition that $\theta_{B} = 0$.
Figure 3 depicts a color map showing the convergence characteristics of orbit $\phi_{i}$ with respect to its initial value and the set value of $\gamma\, ( \in[0, 2 \pi])$. Blank (colorless) regions on the map indicate that $\phi_{i}$ is chaotic or periodic there and does not converge.
As seen in this map, (a):$\phi=0$ is an attractor regardless of the setting $\gamma$, and (b): another attractor $\phi \sim1$ appears because of a kind of catastrophe (not a bifurcation) \cite{Arnold} in the neighborhood of our condition of interest ($\gamma/\pi \sim 1$).
We further refine the conditions and investigate the dissipative causality in detail for $\gamma = \pi$ (the condition in which the attractor due to catastrophe is $1$).
Here, to improve our perspective on the behavior of the system, we introduce a {\it pseudo-force} ${\mathfrak F}( \phi )$ and a {\it pseudo-potential} ${\mathfrak V}(\phi)$, which are defined as follows:

\begin{figure}[t]
\centering\includegraphics[width=7cm]{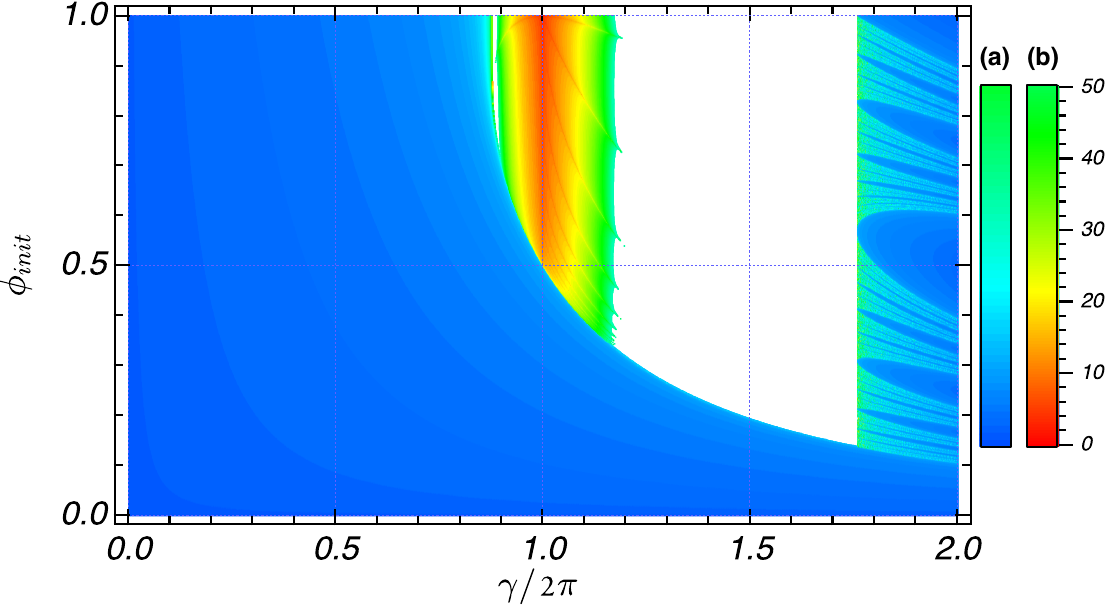}
 \caption{Convergence characteristics (number of steps needed to reach a threshold) of $\phi_{i}$ with respect to initial values $\phi_{\rm init.}$ and $\gamma$ under a condition of $\theta_{B} = 0$. Color gages for convergence to two \it attractors\rm :(a) $\phi_{\rm conv.}=0$ and (b) $\phi_{\rm conv.}\ne 0\, (\sim 1)$.}
 \end{figure}

\begin{eqnarray} 
{\mathfrak F}(\phi) 
& := &
\phi_{i+m}  -  \phi_{i}  \,\, |_{\phi_{j} = \phi}, \,\,\,\,\,\,\,\,\, \phi   \in  [ 0, 1 ]
\nonumber\\
& = &
\frac{1}{2}  \sin( \pi ( \phi - \frac{1}{2} ) )  -  ( \phi - \frac{1}{2} )
\\
- \frac{d}{d\phi} {\mathfrak V}(\phi)
& := &
{\mathfrak F}(\phi),  \,\,\,\,\,\,\,\,\,\,\,\,\,\,\,\,\,\,\,\,\,\,\,\,\,\,\,\,\,\,\,\,\,\,\,\,\, \phi   \in  [ 0, 1 ] 
\nonumber\\
{\mathfrak V}(\phi) 
& = &  
\frac{1}{2\pi}  \cos( \pi ( \phi - \frac{1}{2} ) )  +  \frac{1}{2} ( \phi - \frac{1}{2} )^{2} + c
\end{eqnarray}

In fact, the dissipative causality loses its direct dependence on space-time, which makes it impossible to define the force $F({\bf x} )$ and potential $V({\bf x})$ as in ordinary dynamics.
However, the ${\mathfrak F}(\phi)$ and ${\mathfrak V}(\phi)$ introduced here for convenience are useful to understand the behavior of the trajectory by analogy with ordinary dynamics.

\begin{figure}[t]
\centering\includegraphics[width=7cm]{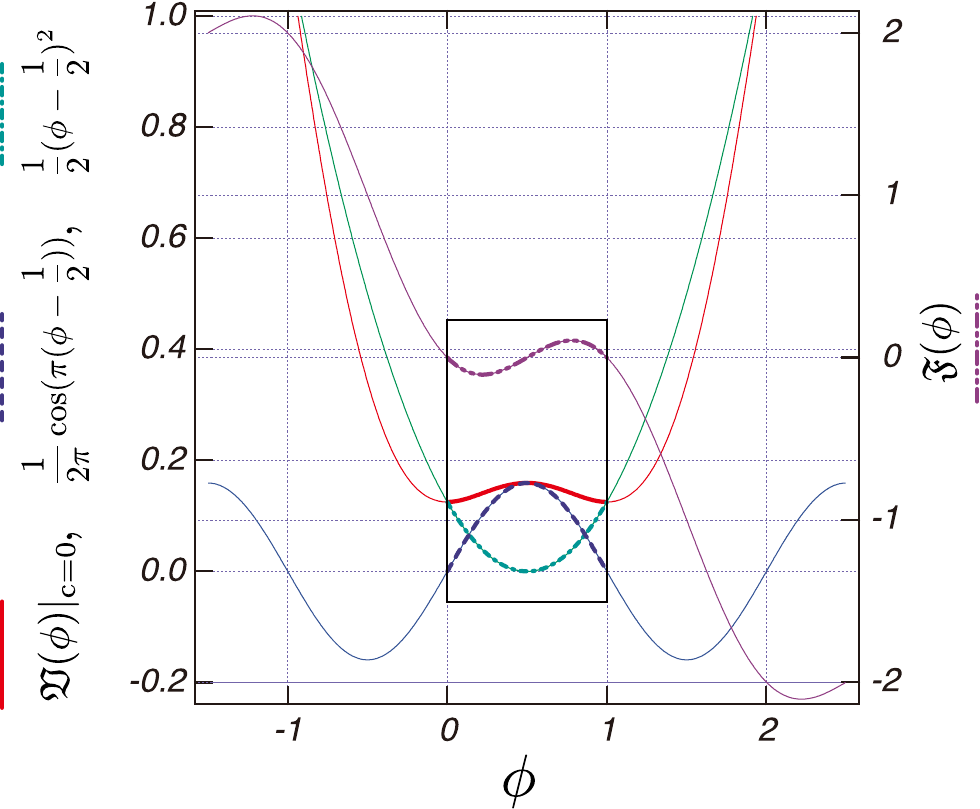}
 \caption{\it Pseudo-force \rm ${\mathfrak F}(\phi)$ and \it pseudo-potential \rm ${\mathfrak V}(\phi)$ defined by Eqs. (5) and (6), respectively ($\phi \in [0, 1]$). Undefined outer region is also drawn mechanically to assist in understanding dissipative causality's behavior under conditions of $\theta_{B} = 0$ and $\gamma = \pi$.}
 \end{figure}

As shown in Fig. 4, ${\mathfrak V}(\phi)$ is similar to the well-known $V({\bf x})$ that was introduced to apply the SSB concept \cite{BCS, Bogoliubov} in statistical physics to examine phase transitions in equilibrium.
This finding strongly suggests that the behavior of dissipative causality under certain conditions can be understood as a kind of SSB.

Previously reported SSBs have generally discussed phase transitions under the assumption of equilibrium states. In these discussions, symmetry and its breaking are treated as changes in the potential $V({\bf x})$ itself due to temperature, etc., using an order parameter that represents the position of stable equilibrium points in the potential $V({\bf x})$. Subsequent chapters of this paper describe a novel approach that achieves the state preceding SSB with restored symmetry not by varying the pseudo potential ${\mathfrak V}(\phi)$ itself via external parameters, but by forcibly transitioning the system state to an unstable fixed point of the (fixed) pseudo-potential. In other words, the new method induces SSB from a state where symmetry is controlled and restored by achieving a system state change that corresponds to changing the position variable on the pseudo potential ${\mathfrak V}(\phi)$. Therefore, to avoid confusion with conventional phase transition discussions, we proceed with the discussion using a pseudo order parameter defined as follows, employing $\phi$, the position variable of the fixed pseudo-potential.
The pseudo order parameter here can be expressed as follows, and at the SBSB, $\tilde{\phi}=0$, which corresponds to the saddle point of the pseudo ${\mathfrak V}(\phi)$ and is also an unstable fixed point (repeller) of the FDCSs.
\begin{eqnarray}
\tilde{\phi}
& = &
\phi - \frac{1}{2}
\end{eqnarray}

We focus on this possibility because the pseudo order parameter's behavior is governed purely by causality; that is, it can be clearly described by Eq.(4).
So, the first thing to check is that it is possible to restore symmetry by controlling the pseudo order parameter.
We considered transitioning the state of FDCSs (pseudo order parameters) en masse to the SBSB by introducing a following mechanism.

\begin{figure}[b]
\centering\includegraphics[width=7cm]{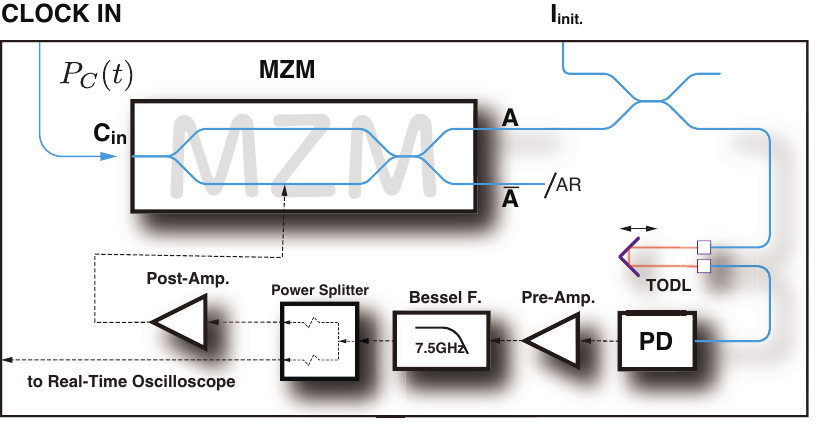}
 \caption{Experimental system for verifying SSB operation in FDCSs. Number of the generated FDCSs $m = 16$. $P_{C}(t)$:whole width of optical pulse train (OPT) 46.08$\mu s$, the OPT cycle 20kHz, repetition frequency of the optical pulses 39.0625MHz, width of the optical pulse 10$ps$. $P_{I}(t)$:whole width of OPT 409.6$ns$, the OPT cycle 20kHz, repetition frequency of the optical pulses 39.0625MHz, width of the optical pulse 10$ps$. Timing when a peak position of an electrical pulse generated from first optical pulse of the $P_{I}(t)$ reaches a MZM was adjusted so that it coincides with timing when the peak position of the 10th optical pulse from the leading edge of $P_{C}(t)$ reaches the MZM.}
 \end{figure}

Mechanism-1:
When the input of $P_{C}(t)$ from ${\bf C_{\rm in}}$ begins and the dissipative causalities are created, a steady state with $\phi_{i}=0$ appears. For this reason, the FDCS does not exhibit the interesting behavior described later. In this paper, we will refer to this initial steady state as a {\it hidden state} (HS) \cite{Comments001}.
Therefore, we introduce an optical input port ${\bf I_{\rm init.}}$ that allows the external input to control the state of the FDCSs, as shown in Fig. 5.
Then, after starting the input of clock pulses $P_{C}(t)$, it is possible to collectively transition the state of the FDCSs from the HS to the SBSB by inputting the following optical initializing pulse train, which has been adjusted for timing, from the ${\bf I_{\rm init.}}$.
\begin{eqnarray}
P_{I}(t)
& = &
\frac{1}{2}
\sum_{j=0}^{m-1}\,
 P^{c}_{n0+j}(t)
\end{eqnarray}

, where $n0$ is an optional position.

\begin{figure}[t]
\centering\includegraphics[width=8cm]{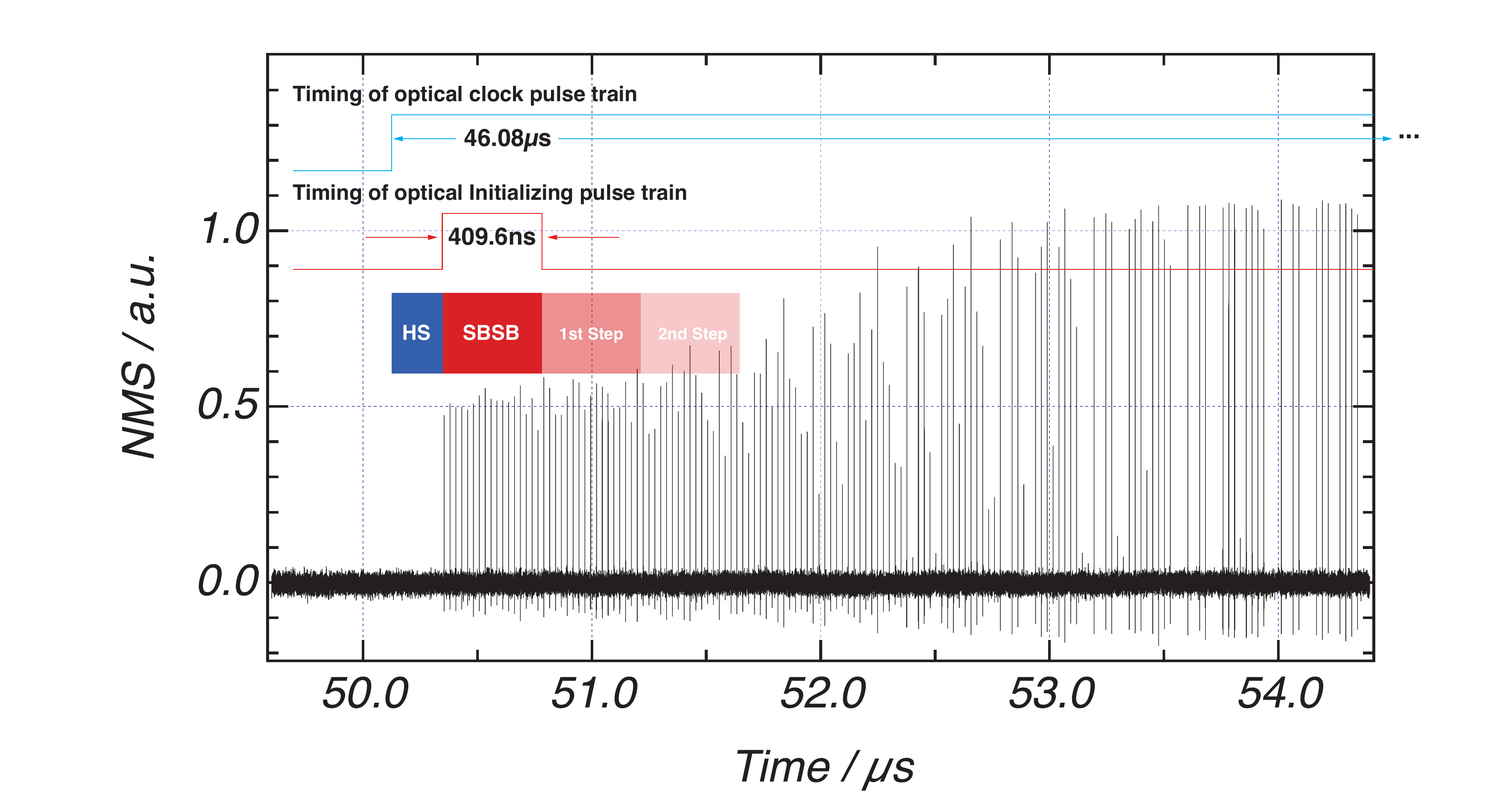}
 \caption{Observed waveform in experimental trial to confirm foundational SSBs in FDCSs (N=16). Based on the input of an optical clock pulse (OCP) train, 16 independent FDCSs in a hidden state are constructed. Furthermore, when an optical initialization pulse train is input, the system transitions to SBSB. While the OCP train input is maintained, the 16 FDCSs step independently of each other due to the dissipative causality of each FDCS.}
 \end{figure}

Using an experimental system with the above-described configuration adjusted so that $m=16$, we will describe an experimental demonstration of the control of state transition to the SBSB, which also serves as a verification of the content discussed and proposed so far.
An optical pulse train with a repetition frequency of 39.0625MHz is prepared by extracting from a series of optical output pulses with a repetition frequency of 5GHz and a pulse width of 10 ps emitted from an active mode lock laser.
Furthermore, in order to enable repeated trials, the OCPs $P_{C}(t)$ with a width of 46.08 $\mu s$ and a cycle of 20 kHz are extracted from this optical pulse train.
$P_{I}(t)$ is a sequence of 16 optical pulses with a pulse width of 10 ps and a repetition rate of 39.0625 GHz, extracted at a cycle of 20kHz, with the delay adjusted to coincide with the start time of the test.
Furthermore, a 1$\times$2 MZM with a modulation bandwidth that is more than sufficient to respond to electrical pulses extended to a pulse width of 200 ps is used as a variable inflow gate.
We tuned the system to the condition $\gamma = \pi$ by adjusting both the input power of the optical clock pulse and the amplification rate of the electrical amplifier.
Figure 6 shows the monitor signal waveform observed in the demonstration trial to confirm the state transition to the SBSB and the subsequent SSB.
Since the normalized optical peak power of the observed pulse signal and the optical transmittance $\phi$ coincide, it is possible to verify the behavior of the pseudo order parameters from this waveform (specifically, the behavior of the normalized peak power).

The successful transition control from the HS to the SBSB, the behavior of the SSB, and the convergence of the pseudo order parameters to the two attractors can be confirmed. These can be said to indicate that the establishment of dissipative causality by the input of $P_{C}(t)$ has been confirmed.
When viewed as a new theme of applying the SSB phenomenon, the success of the above-mentioned experimental verification, combined with the realization of repetitive motion, can be said to be an important milestone.

\section{SSB machine}

In this section, we will further develop the discussion and consider a model in which each SSB creates its elements of {\it a many-body-like system} (MBLS).
And, based on the idea that the attractors (-1/2,1/2) in which the pseudo order parameters are trapped as a result of SSB can be regarded as pseudo-spins (Up/Down), we will narrow down the discussion to an Ising-model-like system, in which the interaction between these pseudo-spins corresponds to an Ising interaction (exchange interaction) in the IMS \cite{Heisenberg, Pauli}.
We will refer to the system that performs simulations using phenomena that have duality with the above-mentioned model as an SSB machine.
When considering giving a concrete expression for the pseudo-spin interaction(PSI), the desired effect cannot be obtained even if the exchange interaction expression is blindly followed, because (I):the system is a discrete dynamical system that obeys the iteration equation, and (II):while the original spin takes a discrete value, the pseudo-spin takes a continuous value.
Therefore, we decided to assign the following properties based on the original interpretation of the exchange interaction energy.
(i-0):In the case of ferrimagnetism: the property of aligning to the same state appears no matter what value the pseudo order parameters take in the range $[-1/2, 1/2]$. (ii-0):In the case of anti-ferrimagnetism: the property of trying to transition to a state where they are in opposite positions to each other appears no matter what value the pseudo order parameters take in the range $[-1/2, 1/2]$.
The PSI that we propose and introduce here is expressed by the following equation, and it is possible to implement it physically by optical interference using Mach-Zehnder interferometers between each element.

\begin{figure}[t]
\centering\includegraphics[width=7cm]{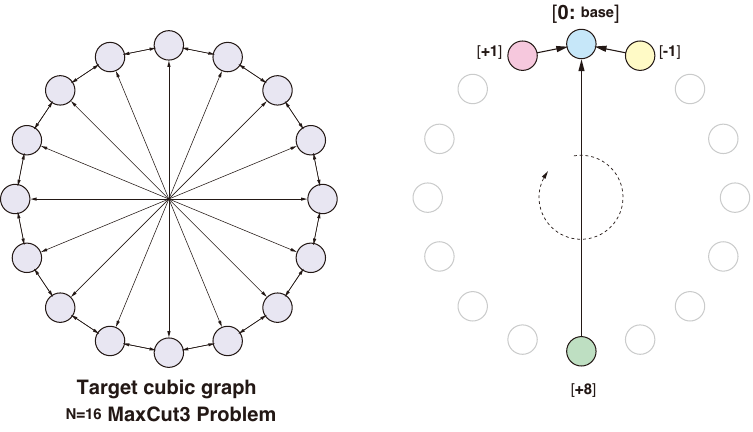}
 \caption{A cubic graph of a target (target COP) for model calculations of SSBM with ${\cal J}_{i:k}^{\rm AF}=1/4\, (k=i-8, i-1, i+1)$ and ${\cal J}_{i:k}^{\rm F}=0$. This cubic graph represents NP-hard instances, equivalent to MaxCut3 (N=16) problem\cite{Takata}.}
 \end{figure}

\begin{eqnarray}
\phi_{i+m}
& = &
\sin^{2}(\frac{\pi}{2} \,|\sqrt{\phi_{i}} -  {\cal Q}_{i} |^{2} ) \\
{\cal Q}_{i}
& = &  
{\cal Q}_{i}^{\rm F} + {\cal Q}_{i}^{\rm AF} \\
{\cal Q}_{i}^{\rm F}
& = &  
- \sum^{i \ne k} {\cal J}_{i:k}^{\rm F}( \sqrt{\phi_{i}}  -  \sqrt{\phi_{i+k}})  \\
{\cal Q}_{i}^{\rm AF}
& = &  
- \sum^{i \ne k} {\cal J}_{i:k}^{\rm AF}( \sqrt{\phi_{i}}  -  \sqrt{\overline{\phi_{i+k}}})
\end{eqnarray}
\begin{eqnarray}
 {\cal J}_{i:k}^{\rm F} = \left\{ \begin{array}{ll}
  {\cal J}_{i:k} & \mbox{if} \mbox{\footnotesize \rm \,\,\,\,\,\,\,\,\,Ferro.\, type}\\
                                 0 & \mbox{if}\,\,\, \mbox{\footnotesize \rm  Anti-Ferro.\, type} \end{array} \right. \\
 {\cal J}_{i:k}^{\rm AF} = \left\{ \begin{array}{ll}  0 & \mbox{if}\,\,\, \mbox{\footnotesize \rm  \,\,\,\,\,\,\,\,\,Ferro.\, type} \\
                                  {\cal J}_{i:k} & \mbox{if} \,\,\,   \mbox{\footnotesize \rm  Anti-Ferro.\, type}  \end{array} \right. 
\end{eqnarray}
where $k \in I, -m/2 \le k < m/2$.  Note that $\overline{\phi}$ is the transmittance to the port $\overline{\bf A}$ and is also the complement of $\phi$ ($\overline{\phi} = 1 - \phi$).

The point to note here is that these PSIs act in concert with the convergence of pseudo order parameters to attractors and that the convergence characteristics are a property of the dynamics themselves.
As a result, it is expected that (i-1):there will be an effect that causes them to transition to the same attractor, or (ii-1):there will be an effect that causes each of them to transition to a different attractor\cite{Comments002}.
Therefore, the final state that the SSBM will reach is thought to be consistent with the stable state (or most stable state) of the IMS.
To verify this prediction, we performed numerical simulations of the SSBM corresponding to the MaxCut3 problem (see Fig. 7), which is characterized by a cubic graph with ${\cal J}_{i:k}^{\rm AF}=1/4\, (k=i-8, i-1, i+1)$ and ${\cal J}_{i:k}^{\rm F}=0$, one of the COPs.
This COP has been taken up in other physically implemented Ising machine studies for the validation of effectiveness \cite{Takata} and is taken up here as a benchmark.
We performed 1,000 numerical simulations using different Gaussian random numbers as the initial fluctuations that trigger SSB to investigate the behavior of the SSBM.
Similarly to the results of the SSB demonstration experiment in the FDCS, in the SSBM, it was confirmed that the system set to the SBSB $(\phi = 1/2)$ as the initial state exhibits a behavior that transitions to one of the two attractors (Fig. 8(a)). However, at the same time, the system selectively transitions to a stable state (optimal state) corresponding to the set pseudo-spin interaction. Figure 8(b) shows the relationship between the probability that the combination of pseudo-spins of the IMLS appearing by the SSBM becomes one of the 16 degenerate most stable state states and the magnitude of the initial fluctuation.
Here, Ising energy $E_{\rm Ising}(\tilde{\mathbf \phi})$ and cut number $C(\tilde{\mathbf \phi})$ were evaluated using pseudo-spins (values of pseudo order parameter in the final state) obtained from the simulation, in accordance with the evaluation method in Ising machines, using the following equations.
\begin{figure}[t]
\centering\includegraphics[width=8cm]{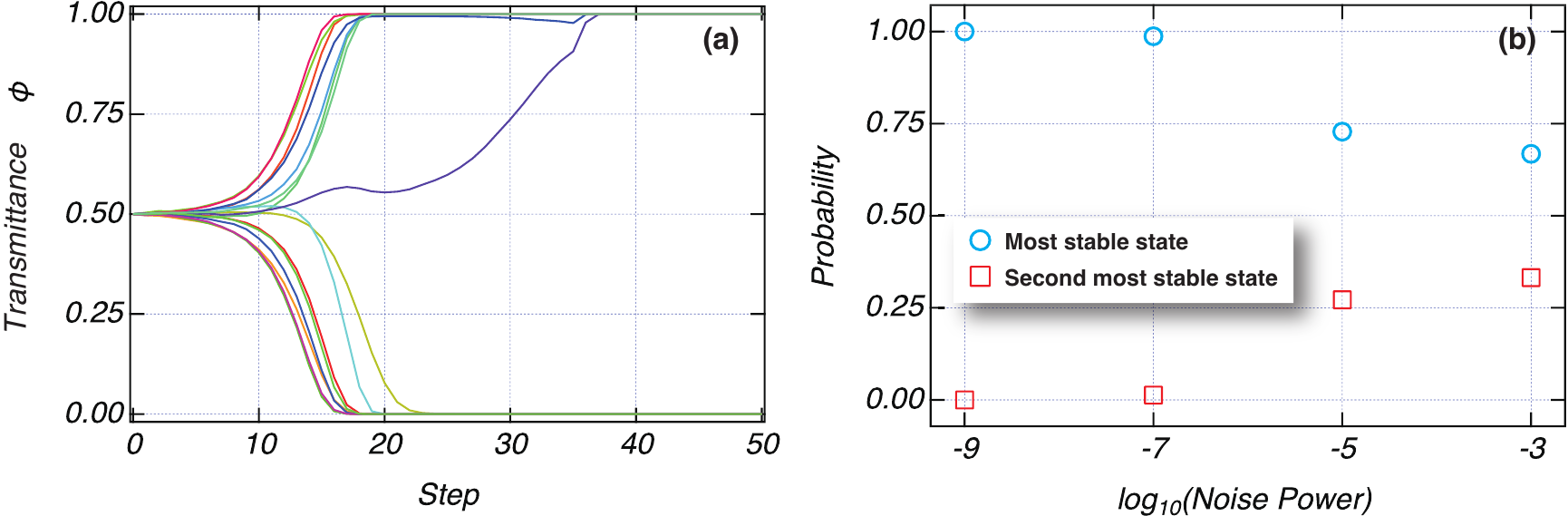}
 \caption{Estimated results based on numerical simulation of SSBM characteristics (MaxCut3, $m=16$, Number of traials:1,000). (a):Transition of 16 transmittance values in SSBM. Similar to the observation results of the SSB phenomenon in the FDCS (Fig. 6), all $\phi_{i}$ set in SBSB are attracted to one of the two attractors. Furthermore, due to the effect of pseudo-spin interactions, a transition to the stable state in the corresponding Ising model occurs simultaneously. (b):Relationship between the probability of SSBM finding a stable state and noise power. Numerical simulation results show that SSBM can find either the most stable state or the second most stable state with 100\% probability, and that when noise power is at or below the $10^{-9}$ level, the most stable state can be found with 100\% probability.}
 \end{figure}
\begin{eqnarray}
E_{\rm Ising}(\tilde{\mathbf \phi})
& = &
- \frac{1}{2}\sum_{i,j}  {\cal J}_{i:j} (2 \tilde{\phi}_{i})(2  \tilde{\phi}_{j})\\
C(\tilde{\mathbf \phi})
& = &
- \frac{1}{2}\sum_{i,j}  {\cal J}_{i:j} \frac{1-(2 \tilde{\phi}_{i})(2  \tilde{\phi}_{j})
}{2}
\end{eqnarray}

These simulation results strongly suggest that the IMLS states appearing in the SSBM correspond to the energy-stable states in the IMS.
In addition, the expected characteristics of the SSBM are comparable to those of previously reported physical implementation type Ising machines \cite{Yamamoto, Takata}, indicating that the SSBM is a promising solver for COPs.

However, in configurations that realize all individual PSIs presented here via optical interference, as the number of PSIs increases, the optical power of each individual PSI must be reduced, making the SN ratio of the optical pulses the root cause of a severe scalability problem. Similarly, for CIM, a solver also using optical pulses, no experimental reports exist of configurations achieving Ising interactions solely through optical interference, even for problems where all spins are coupled. This is because CIM faces challenges similar to those of SSBM. This scalability problem is expected to be solvable by following the example of CIM, which succeeded in experimental application to the fully-connected problem with N=100000 \cite{Honjo}. This scalability issue is expected to be resolved by following the example of CIM, which successfully applied to the N=100,000 fully connected problem. There, the problem is addressed using a method termed a measurement and feedback scheme: optically encoded signals corresponding to spins are converted to electrical signals, then interactions are computed digitally using an FPGA, and finally the computed interactions are feedback as optical signals via electro-optical conversion.

\section{Discussion}

Here, we will point out some of the features of SSBM that should be noted.
First of all, the elements of IMLS (pseudo-spin) have lost their direct dependence on space-time, and they are not even particles to begin with. Therefore, they cannot be treated using important and successful conventional theories such as gauge theory, Lagrangian theory, and Hamiltonian theory, and there is no point in trying to do so.
On the other hand, for models that take into account spatial restrictions such as the placement of pseudo-spins on lattice points, it is possible to utilize the ${\cal J}_{i:k}$ setting degrees of freedom in a convenient manner, following the Ising model.
In the limit where the fluctuation becomes zero, after the collective transition to SBSB (see Mechanism-1), all elements continue to remain in the repeller, and the causal inversion symmetry (corresponding to time-reversal symmetry) and exchange symmetry between elements are also restored.
It is easy to confirm from Eqs.(9) and (10) that this conclusion remains unchanged even when the PSI is taken into account and that it is a characteristic of IMLS.
To put it another way, while dissipative causality is certainly maintained, there is nothing that can make the elements of IMSL relative to each other (the elements are transparent to each other) and even causal inversion symmetry is restored; so to speak, it is {\it nothing} (nothing state (NS)).
In other words, in this model, we should consider that the pseudo-spin is not something that is a priori.
However, because fluctuations exist, symmetry spontaneously breaks down, and pseudo-spin and interrelations (the PSI amplitude) appear in a complementary manner.
The elements gradually become more visible to each other, and eventually a situation emerges in which they can be recognized as a kind of spin.
From the above, we can conclude that this system can be understood as a system in which the IMLS emerges from the NS.
From the perspective of categorizing SSBs, it can also be said that the occurrence of Type-II, -III, and -I SSBs in a layered and complementary manner is a novel and very interesting phenomenon.
Furthermore, the fact that causality plays a vital role is also common to theoretical research on models in which space and time are created from premises without a space-time background \cite{Henson, Sorkin, Ambjorn, Tsuchiya}, and it is also interesting that both of these are concerned with creation phenomena.

Let us now discuss the correspondence found between IMS and IMLS.
Needless to say, spin systems such as the IMS and the Hubbard model are research topics that should be treated in quantum theory.
In these research areas, classical computer simulations become practically difficult as the number of spins in the system increases, so research is being carried out to solve this problem using quantum simulations\cite{Feynman, Jaksch, Greiner}, except when the problem can be solved analytically under special conditions.
In recent years, there have also been active attempts to investigate the behavior of macroscopic physical quantities near the quantum critical point by applying the AdS/CFT correspondence \cite{Maldacena} methodology as a new approach that can avoid the difficulty of realizing quantum simulation systems \cite{Hartnoll, Matsumoto}.
Other examples of research that focuses on duality include the proposal and study of indirect empirical experimental models for difficult-to-observe phenomena such as Hawking radiation\cite{Hawking, Unruh, Kovtun, Katayama}.
These attempts to overcome the difficulties of simulation or observation through approaches based on duality between theoretical/mathematical models or phenomena have attracted attention because they are expected to play an important role in providing a perspective for research that faces difficulties and in advancing such research, although they must compromise on mathematical rigor. The {\it phenomenon} discussed in this paper in SSBM does not have a duality with IMS, but is a {\it phenomenon} that has a duality with the model in which {\it pseudo-spins expressed by continuous pseudo order parameters} are created by the SSB.

In this paper, the first attempt is made to deal with a model that can discuss {\it the creation process and the states that appear at the time of creation}, whereas the discussion of the creation of elementary particles by SSB has conventionally focused on the pre- and post-creation stages \cite{Nambu, Higgs}.
In addition, in this new model, it is expected that the state that finally converges in the IMLS that appears in the SSB corresponds to the stable or most stable state in the IMS, and it was verified that the numerical simulation of the system taken as a benchmark produced the expected results.
Based on the above discussion, we see the SSBM discussed in this paper as one of the studies that focus on the correspondence relationship through duality (an attempt to overcome the difficulty of simulation) - albeit indirectly.

\section{Conclusion}

The theory of dissipative causality created in unique dissipative systems is discussed in detail.
Then, we theoretically elucidated and experimentally demonstrated that this dissipative causality can lead to the development of a new type of SSB phenomenon under certain special conditions.
Furthermore, we proposed a new solver for COP using this new SSB phenomenon, namely SSBM, and confirmed its effectiveness through numerical simulation.
SSBM can be considered a type of research that overcomes simulation difficulties by focusing on correspondences through duality.
The experimental demonstration of SSBM is one of the main issues that needs to be addressed in the future.

\section{Acknowledgment}
The author thanks A. Sugiyama,  M. Tsuda, S. Oka, H. Takenouchi, T. Komatsu, T. Sawada, O. Kagami, S. Mutoh, T. Haga, and A. Okada for their support.


\begin{thebibliography}{40}
\bibitem{Comments003}  
If we focus solely on the two points that the system behavior is described by a iterative equation and that a MZM is used, there exists another example that differs from the one discussed in this paper. However, that example is a system where data processing by a digital data processor dominates, and it integrates the analog response of the MZM into the processing flow using AD/DA converters.Furthermore, the focus of interest in that system lies in investigating the mathematical behavior of the iterative equations themselves and developing a computational experimental system capable of providing high-speed calculation methods, whereas the research in this paper fundamentally differs in that it places its primary interest in special phenomena that can be described by iterative equations.
\bibitem{Comments001} 
In the conditions we are focusing on ($\theta_{B}=0$), there is no change in the $\phi_{i}$ before and after the input of $P_{C}(t)$ in the region shown in Figure 3 ($\gamma / \pi = [0, 2]$), but the dissipative causality is established after the input of $P_{C}(t)$.
Furthermore, when $\gamma / \pi = 1$, the input of $P_{C}(t)$ can be considered to result in the emergence of a state in which spontaneous symmetry breaking has been completed.
\bibitem{Comments002} 
In order to improve the outlook on the problem, we focused on the two-body problem and described it. If we extend it to a general many-body problem, it will show complex behavior.
\end{thebibliography}
\end{document}